\documentclass[proceedings]{stacs}
\stacsheading{2010}{227-238}{Nancy, France}
\firstpageno{227}

\usepackage{amsmath}
\usepackage{amssymb}
\usepackage{graphics}
\usepackage{epic}
\usepackage{eepic}
\usepackage{color}
\usepackage{stmaryrd}
\usepackage{algorithm}
\usepackage{algorithmic}
\usepackage{latexsym}

\newlength{\g}
\newcommand{\Pe}{{\rm P}}
\newcommand{\NP}{{\rm NP}}

\newcommand{\LogCFL}{{\rm LogCFL}}
\newcommand{\Log}{{\rm L}}

\newcommand{\ACZero}{\rm AC$^0$}
\newcommand{\TCone}{\rm TC$^1$}

\begin{document}
\sloppy
%
%
%

\title[Restricted Space Algorithms for Isomorphism on  Bounded Treewidth Graphs]{Restricted Space Algorithms for Isomorphism on  Bounded Treewidth Graphs}

\author[ref1]{B.\ Das}{Bireswar Das}
\address[ref1]{Institute of Mathematical Sciences, Chennai, India}
\email{bireswar@imsc.res.in}

\author[ref2]{J.\ Tor{\'a}n}{Jacobo Tor{\'a}n}
\address[ref2]{Institut f\"{u}r Theoretische Informatik, Universit\"{a}t Ulm, 89069 Ulm, Germany}
\email{jacobo.toran@uni-ulm.de}

\author[ref3]{F.\ Wagner}{Fabian Wagner}
\address[ref3]{Institut\ f\"{u}r Theoretische Informatik, Universit\"{a}t Ulm, 89069 Ulm, Germany}
\email{fabian.wagner@uni-ulm.de}

\thanks{Supported by DFG grants TO 200/2-2.}

\subjclass{Complexity Theory, Graph Algorithms}

\begin{abstract}
The Graph Isomorphism problem restricted to graphs of bounded treewidth or bounded tree distance width 
are known to be solvable in polynomial time~\cite{Bo90},\cite{YBFT}.
We give restricted space algorithms for these problems proving  the following results:

\begin{itemize}
\item Isomorphism for bounded tree distance width graphs is in  \Log\ and thus complete for the class. We also show that for this kind of graphs a canon 
can be computed within logspace.
\item For bounded treewidth graphs, when both input graphs are given together 
with a tree decomposition, 
the problem   of whether there is an isomorphism which respects the decompositions 
(i.e.\  considering only  isomorphisms mapping bags in one decomposition 
blockwise onto bags in the other decomposition)
is in \Log.
\item For bounded treewidth graphs, when one of the input graphs is given with a tree decomposition
the isomorphism problem is in \LogCFL.
\item As a corollary the isomorphism problem for bounded treewidth graphs is in \LogCFL. This improves the 
known \TCone\  upper bound for the problem given by  Grohe and Verbitsky~\cite{GV06}.  
\end{itemize}
\end{abstract}

\maketitle


\section{Introduction}

The Graph Isomorphism problem consists in deciding whether two given graphs are isomorphic, or in other 
words, whether there exists  a bijection between the vertices of both graphs preserving the edge 
relation. Graph Isomorphism  is a well studied problem in \NP\ because of its many applications and also 
because it is one of the few natural problems in this class not known to be solvable in polynomial time 
nor known to be \NP -complete. 
Although for the case of general graphs 
no efficient algorithm for the problem is known, the situation is much better
when certain parameters in the input graphs are bounded by a constant. For example the isomorphism problem for 
graphs of
bounded degree~\cite{Luks1}, bounded genus~\cite{Mi80}, bounded color classes~\cite{Luks2}, or bounded treewidth~\cite{Bo90} 
is known to be in \Pe. Recently some of these upper bounds 
have been improved with the development 
of space efficient techniques,
most notably  Reingold's
deterministic logspace algorithm for connectivity in undirected graphs~\cite{Re08}.
In some cases  logspace algorithms have been
obtained. For example graph isomorphism for trees~\cite{Lin92}, planar graphs~\cite{DLNTW} or $k$-trees~\cite{KK09}.
In other cases the problem has been classified in some other small complexity classes below \Pe.
The isomorphism problem
for graphs of bounded treewidth is known to be in \TCone~\cite{GV06}  
and the problem restricted to graphs of bounded color classes is known to be in the $\#$\Log\ hierarchy~\cite{AKV}.

In this paper we address the question of whether  the isomorphism problem restricted to graphs of bounded treewidth and bounded tree distance 
width can be solved in logspace. Intuitively speaking, the treewidth of a graph measures how much it differs from a tree.
This concept has been used very successfully in algorithmics and fixed-parameter tractability (see e.g.~\cite{Bo96,BK07}). 
For many complex problems, efficient algorithms have been found for the cases
when the input structures have bounded 
treewidth. As mentioned above  Bodlaender showed in~\cite{Bo90} that  Graph Isomorphism can be solved in 
polynomial time when restricted to graphs of bounded treewidth.  
More recently Grohe and Verbitsky~\cite{GV06} improved this upper bound
to \TCone.
In this paper we improve this result  showing that the isomorphism problem for  bounded treewidth graphs  
lies in \LogCFL, the class of
problems logarithmic space reducible to a context free language. \LogCFL\ can be alternatively characterized as the class of problems computable 
by a uniform family of polynomial size and logarithmic depth circuits with bounded AND and unbounded OR gates, and is therefore a subclass of \TCone.  
\LogCFL\  is also the best known upper bound for
computing a tree decomposition of bounded treewidth graphs~\cite{Wa94,GLS}, which is one bottleneck
in our  isomorphism algorithm. We prove that if 
tree decompositions of both graphs are given as part of the input, the 
question of whether there is an isomorphism respecting the vertex partition defined by the  decompositions can
be solved in logarithmic space. Our proof techniques are based on methods from 
recent isomorphism results~\cite{DLNTW,DNTW}
and are very different from those in~\cite{GV06}.

The notion of tree distance width, a stronger version of the treewidth concept, was introduced
in~\cite{YBFT}. There it is shown  
that for graphs with  bounded tree distance width the isomorphism problem is fixed parameter tractable, something
that is not known to hold for the more general class of bounded treewidth graphs. 
We prove that for graphs of bounded tree distance width it is possible to 
obtain a tree distance decomposition within logspace. Using this result
we show  that graph isomorphism for bounded tree distance 
width graphs can also be solved in logarithmic space. 
 Since it is known that the question  is also hard for the class \Log\ under \ACZero\ reductions~\cite{JKMT}, this 
exactly 
characterizes the complexity 
of the problem. 
We show that in fact a canon for graphs of bounded tree distance width, i.e.\ a fixed 
representative of the 
isomorphism equivalence class, can be computed in logspace. 
Due to space reasons, some proofs are omitted and will be provided in the  
full version of the paper.


\section{Preliminaries}

We introduce the complexity classes used in this paper.
\Log\ is the class
of decision problems computable by deterministic  logarithmic space 
Turing machines.
\LogCFL\ consists of all  decision problems that can be Turing reduced in 
logarithmic space to a context free language. There are several alternative
more intuitive characterizations of \LogCFL. Problems in this class can be
computed by uniform families of polynomial size and logarithmic depth circuits
over bounded fan-in AND gates and unbounded fan-in OR gates. We will also 
use the characterization of \LogCFL\ 
as the class of decisional problems computable by 
non-deterministic auxiliary pushdown machines (NAuxPDA). These are 
Turing machines with a logarithmic space work tape, an additional pushdown 
and a polynomial time bound ~\cite{Su77}. The class \TCone\ contains the 
problems computable by uniform families of polynomial size and logarithmic 
depth threshold  circuits.
The known relationships among these classes are:\\
\centerline{\Log\ $\subseteq$ \LogCFL\ $\subseteq$ \TCone.}

In this paper we consider undirected simple graphs with no self loops.
For a graph $G=(V,E)$ and two vertices $u,v\in V$, $d_G(u,v)$ denotes the distance between $u$ and
$v$ in $G$ (number of edges in the shortest path between $u$ and $v$ in $G$). For a set $S\subseteq
V$, and a vertex $u\in V$, $d_G(S,u)$ denotes min$_{v\in S}d_G(v,u)$. $\Gamma(S)$ denotes the set of 
neighbors of $S$ in $G$. In a connected graph $G$, a separating set is a set of vertices such that 
deleting the vertices in $S$ (and the edges connected to them) produces more than one connected component.
For $G=(V,E)$ and two disjoint subsets $U,W$ of $V$ 
we use the following notion for an \emph{induced bipartite subgraph} 
$B_G[U,W]$ of $G$ on vertex set $U \cup W$ with edge set $\{ \{u,w\} \in E \mid u \in U, w \in W \}$.
Let $G[U]$ be the \emph{induced subgraph} of $G$ on vertex set $V \setminus U$.

A \emph{tree decomposition} of a graph $G=(V,E)$ is a pair 
$(\{X_i\ |\ i\in I\},T=(I,F))$, 
where $\{X_i\ |\ i \in I \}$ is a collection of  subsets of $V$ called bags, 
and $T$ is a tree with node set $I$ and edge set $F$,
satisfying the following properties:

\begin{itemize}
\item[$i)$] $\bigcup_{i\in I} X_i=V$ 
\item[$ii)$] for each $\{u,v\}\in E$, there is an $i\in I$ with $u,v\in X_i$
and 
\item[$iii)$]  for each $v\in V$, the set of nodes $\{i\ |\ v\in X_i\}$ forms a 
subtree of $T$.

\end{itemize}

The \emph{width} of a tree  decomposition of $G$, 
is defined as $\max\{|X_i|\ |\ i\in I\}-1$.
The \emph{treewidth} of $G$  is
the minimum width over all tree decompositions of $G$.

A \emph{tree distance decomposition} of a graph $G=(V,E)$ is a triple $(\{X_i\ |\ i 
\in I \},T=(I,F),r)$, 
where $\{X_i\ |\ i \in I \}$ is a collection of  subsets of $V$ called \emph{bags}, 
$X_r=S$ a set of vertices and $T$ is a tree with node set $I$,
edge set $F$ and root $r$, satisfying: 

\begin{itemize}
\item[$i)$] $\bigcup_{i\in I} X_i=V$ and for all $i\not=j, X_i\cap X_j=\emptyset$ 
\item[$ii)$] for each $v\in V$, if $v\in X_i$ then $d_G(X_r,v)=d_T(r,i)$ 
and

\item[$iii)$]  for each $\{u,v\}\in E(G)$, there are $i,j\in I$ with $u\in X_i, v\in X_j$ and $i=j$
	or $\{i,j\}\in F$ (for every edge in $G$ its two endpoints belong 
	to the same or to adjacent bags in $T$).
\end{itemize}

Let $D = (\{X_i\ |\ i \in I \},T=(I,F),r)$ be a tree distance decomposition of $G$.
$X_r$ is the \emph{root bag} of $D$.
The \emph{width} of $D$
is the maximum number of elements of a bag $X_i$.
The \emph{tree distance width} of $G$  is
the minimum width over all tree distance decompositions of $G$.

The tree distance decomposition $D$ is called \emph{minimal} 
if for each $i\in I$, the set of vertices in the bags with labels 
in the subtree rooted at $i$ in $T$ induce a connected subgraph in $G$. 
In~\cite{YBFT} it is shown that for every 
root set $S\subseteq V$ there is a unique minimal tree distance decomposition of $G$ with root set $S$. 
The width of such a decomposition is minimal among the tree distance decompositions of $G$ with 
root set $S$.

An isomorphism from $G$ onto $H$ \emph{respects} their tree (distance) decompositions $D,D'$
if vertices in a bag of $D$ in $G$ are mapped blockwise onto vertices in a bag of $D'$ in $H$.
Not every isomorphism has this property.

$Sym(V)$ is the \emph{symmetric group} on a set~$V$.

\section{Graphs of bounded tree distance width}
\subsection{Tree distance decomposition in L}\label{treedistdec}

We describe an algorithm that on input a graph $G$ and a subset $S\subset V$ produces 
the minimal tree distance decomposition $D=(\{X_i\ |\ i \in I \},T=(I,F),r)$ of $G$ with root set 
$X_r=S$. The algorithm works within space
$c\cdot k\log n$ for some constant $c$, where $k$ is the width of the minimal tree distance 
decomposition of $G$ with root set $S$. The output of the algorithm is a sequence of 
strings of the form
( bag label, bag depth, $v_{i_1},v_{i_2},\dots,v_{i_l}$), indicating the number of the bag, 
the distance of its elements to $S$ and the list of the elements in the bag.

The algorithm basically performs a depth first traversal of the  tree $T$ in the decomposition
while constructing it. Starting at $S$ the algorithm uses three functions for traversing $T$.
These functions perform queries to 
a logspace subroutine  computing reachability~\cite{Re08}. 

\bigskip
{\bf Parent}$(X_i)$: On input the elements of a bag $X_i$ the function returns the elements of the
parent bag in $T$. These are the vertices $v\in V$   with the following two properties:
$v\in \Gamma(X_i)\setminus X_i$ and $v$ is reachable from $S$ in $G\setminus X_i$. For a vertex $v$ 
these two properties can be tested
in space $O(\log n)$ by an algorithm with input $G,S$ and $X_i$. In order to find all the 
vertices in the parent set, the algorithm searches through all the vertices in $V$.

\bigskip
{\bf First Child}$(X_i)$: This function returns the elements of the first child of $i$ in $T$. This
is the child with the vertex $v_j\in V$ with the smallest index $j$. $v_j$ satisfies
that $v_j\in \Gamma(X_i)\setminus X_i$ and that $v_j$ is not reachable from $S$ in $G\setminus X_i$. 
It can be found
cycling in order through the vertices of $G$ until the first one satisfying the properties is 
found. The other elements $w \in X_i$ must satisfy the same two properties as 
$v_j$ and additionally, they must be in the same connected component in $G\setminus X_i$ 
where $v_j$ is contained. 
In case $X_i$ does not have any children, the function outputs some special
symbol.

\bigskip
{\bf Next Sibling}$(X_i)$: This function first computes $X_p:=$Parent$(X_i)$ and then searches for the
child of $p$ in $T$ next to $X_i$. 
Let $v_i$ be the vertex with the smallest label in $X_i$.
This is done similarly as the computation of First Child. 
The next sibling is the bag containing the unique vertex $v_j$ 
with the following properties:
$v_j$ is  the vertex with the smallest label in this bag,
$label(v_j) > label(v_i)$ and there is no other bag which has a vertex with a label $>v_i$ and $<v_j$.
The vertex $v_j$ is not reachable from $S$ in $G\setminus X_p$.
The other elements in the bag are the vertices satisfying these properties and which are in the same connected
component of $G\setminus X_p$ where $v_j$ is contained.

\bigskip

With these  three  functions the algorithm
performs a depth-first traversal of $T$. It only
needs to remember the initial bag $X_0=S$ which is part of the input, and the elements of 
the current bag. On a bag $X_i$ it searches for its first child. If it does not
exist then it searches for the next sibling. When there are no further siblings the next move goes
up in the tree $T$. The algorithm finishes when it returns to $S$. 
It also keeps two counters in order to be able to output the number and depth of the 
bags. The three mentioned functions only need to keep at most two bags ($X_i$ and its father) 
in memory, and work in logarithmic space. On input a graph $G$ with $n$ vertices, and a root set $S$,
the  space used by the algorithm is therfore bounded by $c\cdot k\log n$, for a constant $c$,
and $k$ being the minimum width of a tree distance decomposition of $G$ with root set $S$.
When considering how the three functions are defined 
it is clear that the algorithm constructs a tree distance decomposition
with root set $S$. 
Also they make sure that for each $i$ the subgraph
induced by the vertices of the bags in the 
subtree rooted at $i$ is connected thus producing a minimal decomposition. 
As observed in~\cite{YBFT}, this is the
unique minimal tree distance decomposition of $G$ with root set $S$. 


\subsection{Isomorphism Algorithm for Bounded Tree Distance Width Graphs}

For our isomorphism algorithm we use a tree called the \emph{augmented
 tree} which is based on the underlying tree of a minimal tree
distance decomposition. This augmented tree, apart from the bags,
contains information about the separating sets which separate bags.

\begin{definition}
Let $G$ be a bounded tree distance width graph with a minimal tree distance
decomposition $D=(\{X_i\ |\ i \in I \},T=(I,F),r)$.
The \emph{augmented tree} $\mathcal{T}_{(G,D)} = (I_{(G,D)},F_{(G,D)},r)$ 
corresponding to $G$ and $D$ is
a tree  defined as follows:

\begin{itemize}
\item The set of nodes of $\mathcal{T}_{(G,D)}$ is $I_{(G,D)}$ which contains
	two kinds of nodes, namely $I_{(G,D)} = I \cup J$.
	Those in $I$ form the set of \emph{bag nodes} in $D$,
	and those in $J$ the \emph{separating set nodes}.
	For each 
	bag node $a\in I$ and each child $b$ of $a$ in $T$ we consider the set
	$X_a \cap \Gamma(X_b)$, i.e.\ the \emph{minimum separating set} 
	in $X_a$ which separates $X_b$ from the root bag $X_r$ in $G$.
	Let $M_{s^a_1},\dots,M_{s^a_{l(a)}}$ be the set of all minimum separating sets 
	in $X_a$, free of duplicates. There are 
	nodes for these sets $s^a_1,\dots,s^a_{l(a)}$, the separating set nodes.
	We define $J = \bigcup_{a \in I} \{ s^a_1,\dots,s^a_{l(a)}\}$. 
	The node $r \in I$ is the root in $\mathcal{T}_{(G,D)}$.
\item In $F_{(G,D)}$ there are edges between bag nodes $a \in I$ and 
	the separating set nodes
	$s^a_1,\dots,s^a_{l(a)} \in J$ (edges between bag nodes and their children in the augmented tree).
	There are also   edges between nodes $b \in I$ and $s^a_j$ if $M_{s^a_j}$ 
	is the minimum separating set in $X_a$ which separates $X_b$ from $X_r$
	(edges between bag nodes and their parents).
\end{itemize}
\end{definition}

To simplify notation, we later say for example that
$s_1,\dots,s_l$ are the children of a bag node $a$
if the context is clear.
The odd levels of the augmented tree $T'$ correspond  to bag nodes and
the even levels correspond to  separating set nodes. 

Observe that for each node in the augmented tree, 
we associate a bag to a bag node and a minimum separating set to a separating set node.
Hence, every vertex $v$ in the original graph
occurs in at least one associated component and it
might occur in more than one,
e.g.\ if $v$ is contained in a bag and in a minimum separating set.

Let $T_{(G,D)}$ be an augmented tree of some minimal tree distance
decomposition $D$ of a graph $G$. Let $a$ be a node of $T_{(G,D)}$. The
subtree of $T_{(G,D)}$ rooted at $a$ is denoted by $T_a$.
Note that $T_{(G,D)}=T_{r}$ where $X_r$ is the bag corresponding to the root of
the tree distance decomposition $D$.
We define ${\sf graph}(T_a)$ as the
subgraph of $G$ induced by all the vertices
associated to at least one of the nodes of $T_a$.
The \emph{size} of $T_a$, denoted $|T_a|$
is the number of vertices which occur in at least one component which is 
associated to a node in $T_a$.
Note, $|T_a|$ is polynomially related to $|{\sf graph}(T_a)|$, 
i.e.\ the number of vertices in the corresponding subgraph of $G$.

When given a tree distance decomposition  the augmented tree can be computed in logspace.
Using the result in Section~\ref{treedistdec} we immediately get:

\begin{lemma}
Let $G$ be a graph of bounded tree distance width.
The augmented tree for $G$ can be computed in logspace.
\end{lemma}

\begin{figure}[t]
\begin{center}
\scalebox{0.75}{\input{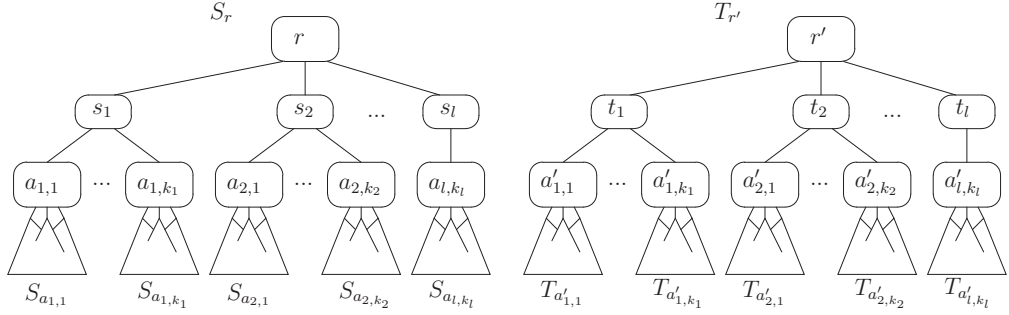}}
\end{center}
\caption{The augmented trees $S_r$ and $T_{r'}$ rooted at bag nodes $r$ and $r'$. 
Node $r$ has separating set nodes $s_1,\dots,s_l$ as children. The children of $s_1$ 
are again bag nodes $a_{1,1},\dots,a_{1,k_1}$. 
$S_{a_{i,j}}$ is the subtree rooted at $a_{i,j}$.
Bag nodes and separating set nodes alternate in the tree.
}
\label{fig:augmTree}
\end{figure}

\paragraph{\bf Isomorphism Order of Augmented Trees.}
We describe an isomorphism order procedure for comparing two 
augmented trees $S_{(G,D)}$ and $T_{(H,D')}$ corresponding to the
graphs $G$ and $H$ and their tree distance decompositions $D$ and $D'$, 
respectively.
This isomorphism order algorithm is an extension of 
the one for trees given by Lindell~\cite{Lin92} and it is different from 
that for planar graphs given by Datta et.al.~\cite{DLNTW}.
The trees $S_{(G,D)}$ and $T_{(H,D')}$ are rooted at bag nodes $r$ and $r'$.
The rooted trees are denoted then $S_r$ and $T_{r'}$ as shown in Figure~\ref{fig:augmTree}.

We will show that 
two graphs of bounded tree distance width are isomorphic if and only if 
for some root nodes $r$ and $r'$ the augmented trees corresponding to the 
minimal tree distance decompositions have the same isomorphism order.

The isomorphism order depends on the order of the vertices in the bags $r$ and $r'$.
Let $X_r$ and $X'_{r'}$ be the corresponding bags in $D$ and $D'$.
We define the sets of mappings $\Theta_{(r,r')} = Sym(X_r) \times Sym(X'_{r'})$.
Let $(\sigma,\sigma')$ be such a mapping,
then the tuples $(G[X_r], \sigma)$ and $(G[X'_{r'}],\sigma')$ 
describe a fixed ordering on the vertices of the induced subgraphs.
If $r$ is not the top-level root of the augmented tree 
then $\Theta_{(r,r')}$ may become restricted to a subset, when going into recursion.
The isomorphism order is defined to be $S_r <_{\tt T} T_{r'}$ if
there exist mappings $(\sigma,\sigma') \in \Theta_{(r,r')}$ such that one of the following holds:

\begin{itemize}
\item[1)] $(G[X_r], \sigma) < (H[X'_{r'}], \sigma' )$ via lexicographical comparison of both ordered subgraphs
\item[2)] $(G[X_r], \sigma) = (H[X'_{r'}], \sigma' )$ but
	$|S_r| < |T_{r'}|$
\item[3)] $(G[X_r], \sigma) = (H[X'_{r'}], \sigma' )$ and $|S_r| = |T_{r'}|$ but
	$\# r < \# r'$ where $\# r$ and $\# r'$ is the number of children of $r$ and $r'$
\item[4)] $(G[X_r], \sigma) = (H[X'_{r'}], \sigma' )$ and $|S_r| = |T_{r'}|$ and $\# r = \# r' = l$ but
	$( S_{s_1},\dots, S_{s_l} ) <_{\tt T} ( T_{t_1},\dots, T_{t_l} )$ where we assume that
	$S_{s_1} \leq_{\tt T}\dots \leq_{\tt T} S_{s_l}$ and $T_{t_1} \leq_{\tt T}\dots \leq_{\tt T} T_{t_l}$ are
	ordered subtrees of $S_r$ and $T_{r'}$, respectively.
	To compute the order between the subtrees $S_{s_i} \leq_{\tt T} T_{t_j}$
	we consider
\begin{description}
	\item[$i$] the lexicographical order 
		of the minimal separating sets ($s_i$ and $t_j$) in $X_r$ and $X'_{r'}$
		according to $\sigma$ and $\sigma'$, as the primary criterion 
		(observe  that the separating sets are subsets of $X_r$ (resp. $X_{r'})$ 
		and are therefore ordered by $\sigma$ and $\sigma'$) and  
	\item[$ii$] pairwise the children $a_{i,i'}$ of $s_i$ and $a'_{j,j'}$ of $t_j$ 
		(for all $i'$ and $j'$ via cross-comparisons) such that
		the induced bipartite graphs $B_G[s_i,a_{i,i'}]$ and $B_H[t_j,a'_{j,j'}]$
		can be \emph{matched} according to $\sigma$ and $\sigma'$ 
		(i.e.\ $\sigma\sigma'^{-1}$ is an isomorphism) and
	\item[$iii$] \emph{recursively} the subtrees rooted at the children of $s_i$ and $t_j$. 
		Note, that these children are again bag nodes. 
		For the cross camparison of bag nodes
		$a_{i,i'}$ and $a'_{j,j'}$ we restrict the set
		$\Theta_{(a_{i,i'},a'_{j,j'})}$ 
		to a subset of $Sym(X_{a_{i,i'}}) \times Sym(X'_{a'_{j,j'}})$.
		Namely,
		$\Theta_{(a_{i,i'},a'_{j,j'})}$ contains the pair 
		$(\phi,\phi') \in Sym(X_{a_{i,i'}}) \times Sym(X'_{a'_{j,j'}})$ if 
		$\phi\phi'^{-1}$ extends the partial isomorphism $\sigma \sigma'^{-1}$ 
		from child $a_{i,i'}$ onto $a'_{j,j'}$ blockwise 
		and which induces an isomorphism from
		$B_G[s_i,a_{i,i'}]$ onto $B_H[t_j,a'_{j,j'}]$.
\end{description}
\end{itemize}

We say that two augmented trees $S_r$ and $T_{r'}$ are \emph{equal according to the isomorphism order}, 
denoted $S_r =_{\tt T} T_{r'}$, if neither $S_r <_{\tt T} T_{r'}$ nor $T_{r'} <_{\tt T} S_r$ holds.

\paragraph{\bf Isomorphism of two subtrees rooted at bag nodes $r$ and $r'$}
We have constant size components associated to the bag nodes. 
A logspace machine can easily run through all the mappings of $X_r$ and $X'_{r'}$ and
record the mappings which gives the minimum isomorphism order. This can be done 
with cross-comparison of trees $(S_r,\sigma)$ and $(T_{r'},\sigma')$ 
with all possible mappings $\sigma,\sigma'$.
Later we will see, that in recursion not all possible mappings for $\sigma$ and $\sigma'$ are considered.
Observe that $|Sym(X_r)| \in O(1)$.

The comparison of $(S_r,\sigma)$ and $(T_{r'},\sigma')$ itself
can be done simply by 
renaming the vertices of $X_r$ and $X'_r$ according to the mappings $\sigma$ and $\sigma'$
and then comparing the ordered sequence of edges lexicographically.
When equality is found then we recursively compute the isomorphism order 
of the subtrees rooted at the children of $r$ and $r'$.

\paragraph{\bf Isomorphism of two subtrees rooted at separating set nodes $s_i$ and $t_j$}

Datta et.al.~\cite{DLNTW} decompose biconnected planar graphs into triconnected components 
and obtain a tree on these components and separating pairs, i.e.\ separating sets of size two.
We have separating sets of arbitrary constant size.

Since $s_i$ and $t_j$ correspond to subgraphs of $X_r$ and $X'_{r'}$, we have an order
for them given by the fixed mappings $\sigma$ and $\sigma'$.
Therefore, 
we can order the children $s_1,\dots,s_l$ and $t_1,\dots,t_l$ according to their occurrence in $X_r$ and $X'_{r'}$
(e.g.\ assume $s_i = (1,2,3,7)$ according to the mapping $\sigma$ and also $s_j = (1,2,4,7)$,
then we get $(s_i,\sigma) <_{\tt T} (s_j,\sigma)$).
Hence, 
when comparing $s_i$ with $t_j$ 
we have to check whether both come on the same position in that order
of $s_1,\dots,s_l$ and $t_1,\dots,t_l$.
If so, then we go to the next level in the tree, to the children of $s_i$ and $t_j$.

Now we have a cross comparison among the children of $s_i$ and the children of $t_j$.
In Steps $4i$, $4ii$ and $4iii$ we partition the children $a_{i,1},\dots, a_{i,l_i}$ of $s_i$ 
and $a'_{j,1},\dots,a'_{j,l_j}$ of $t_j$, respectively,
into isomorphism classes, step by step.

The membership of a child to a class according to Step $4i$ and $4ii$ can be recomputed.
It suffices to keep counters on the work-tape to notice the current class and 
traversing the siblings from left to right.
After these two steps, $a_{i,i'}$ and $a'_{j,j'}$ are in the same class if and only if 
	vertices of $s_i$ and $t_j$ appear lexicographically at the same positions in $\sigma$ and $\sigma'$ and
	the bipartite graphs $B[s_i,a_{i,i'}]$ and $B[t_j,a'_{j,j'}]$ are isomorphic where
	$s_i$ is mapped onto $t_j$ blockwise corresponding to $\sigma \sigma'^{-1}$
	in an isomorphism.
In Step $4iii$ we go into recursion and compare members of one class which 
are rooted at subtrees of the same size.
When going into recursion at $a_{i,i'}$ and $a'_{j,j'}$ 
we consider only those mappings from $(\phi,\phi') \in \Theta_{(a_{i,i'}, a'_{j,j'})}$
which induce an isomorphism $\phi \phi'^{-1}$ from $B[s_i,a_{i,i'}]$ onto $B[t_j,a'_{j,j'}]$.

\paragraph{\bf Correctness of the isomorphism order.}
Both,
the bag nodes and the separating set nodes correspond to subgraphs which are 
basically  separating sets.
A bag separates all its subtrees from the root and the 
separating set nodes refine the bag to separating sets of minimum size.
Hence, 
a partial isomorphism is constructed and extended from each node to its 
child nodes,
traversing the augmented tree (the whole graph, accordingly) in depth first manner.
In the recursion,
the isomorphism between  
the roots of the current subtrees, say $S_r$ and $T_{r'}$,
is partially fixed by the partial isomorphism between their parents.
With an exhaustive search we check every possible remaining isomorphism from $X_r$ onto $X'_{r'}$ 
and go into recursion 
again partially fixing the isomorphism for the subtrees rooted at children of $r$ and $r'$.
By an inductive argument, 
the partial isomorphism described for the augmented tree can be followed simultaneously 
in the original graph and we get:

\begin{theorem}
The graphs $G$ and $H$ of bounded tree distance width are isomorphic 
if and only if
there is a choice of a root bag $r$ and $r'$ producing 
augmented trees $S_r$ and $T_{r'}$
such that $S_r =_{\tt T} T_{r'}$.
The isomorphism order between two augmented trees of $G$ and $H$ 
can be computed in logspace.
\end{theorem}

The proof is based on a careful space analysis at each computational step
building on concepts of the isomorphism order algorithm of Lindell~\cite{Lin92}.
The isomorphism order is the basis for a canonization procedure.
This is shown in a full version of this paper.

\begin{theorem}\label{thm:TreeDistWidthCanoninL}
A graph of bounded tree distance width can be canonized in logspace.
\end{theorem}

\section{Graphs of bounded treewidth}

In this section we consider several isomorphism problems for graphs of 
bounded treewidth. We are  interested in isomorphisms
\emph{respecting} the decompositions (i.e.\ vertices are mapped blockwise from a bag to another bag).
We show first that if the tree decomposition of 
both input graphs is part of the input 
then the isomorphism problem can be decided in \Log.
We also show that if a tree decomposition of only one of the two given 
graphs is 
part of the input, then the isomorphism problem is in \LogCFL. 
It follows that the isomorphism problem for graphs of bounded treewidth is also in \LogCFL.

Assume the decompositions of both input graphs are given.
Let $(G,D),(H,D')$ be two bounded treewidth graphs together with 
tree decompositions $D$ and $D'$, respectively.
We look for an isomorphism between $G$ and $H$ satisfying the condition 
that the images of the vertices in one bag in $D$ belong to the same
bag in $D'$.

We prove that this problem is in  \Log. 
For this we show that given tree decompositions together with designated bags as roots
for $G$ and $H$ the question of whether there is an isomorphism between the graphs mapping root to root 
and respecting the 
decompositions (i.e.\ mapping bags in $G$ blockwise onto bags in $H$) can be reduced 
to the isomorphism problem for graphs
of bounded tree distance decomposition.
We argued in the previous section that this problem belongs to \Log.

\begin{theorem}\label{thm:TreeWIsoToTreeDWIso}
The isomorphism problem for bounded treewidth graphs with given tree decompositions
reduces to
isomorphism for bounded tree distance width graphs
under \ACZero\ many-one reductions.
\end{theorem}

Since bounded tree distance width GI is in \Log, this almost proves the desired result.
To obtain it, we have to  find roots for the tree decompositions.
We fix an arbitrary bag in the one graph
and try all bags from the decomposition of the other graph as roots. We get:

\begin{corollary}\label{cor:TreeWIsoTwoDecompInL}
For every $k\geq 1$ there is a logarithmic space algorithm that, on input a 
pair of graphs together with a tree decompositions of 
width~$k$ for each of them, decides whether there is an isomorphism between
the graphs, respecting the decompositions.
\end{corollary}

\subsection{A~\LogCFL~algorithm for isomorphism} \label{ssec:LogCFLalgForIso}

We consider now the more difficult situation in which  only one of the 
input graphs is given together with a tree decomposition.

\begin{theorem}\label{thm:TreeWIsoOneDecompInLogCFL}
Isomorphism testing for two graphs of bounded treewidth, when  
a tree decomposition for one of them is given, can be done in \LogCFL.
\end{theorem}

\proof	
We describe an algorithm which runs on a non-deterministic auxiliary pushdown 
automaton (NAuxPDA).
Besides a read-only input tape and a finite control, this machine has access to a stack of 
polynomial size
and a~$O( \log n)$ space bounded work-tape.
On the input tape we have two graphs $G,H$ of treewidth $k$ and
a tree decomposition $D=(\{X_i \mid i \in I\}, T=(I,F), r)$ for $G$.
For $j \in I$ we define 
$G_j$ to  be the subgraph of $G$ induced on the vertex set
$\{ v \mid v \in X_i, i \in I$ and~$i=j$ or $i$ a descendant of~$j$ in~$T \}$.
That is, $G_j$ contains the vertices which are separated by the bag $X_j$ from $X_r$ and those in $X_j$.
We define $D_j=(\{X_i, \mid, i \in I_j\}, T_j=(I_j,F_j), j)$ 
as the tree decomposition of $G_j$ 
corresponding to $T_j$, the subtree of $T$ rooted at $j$.
We also consider a way 
to order the children of a node in the tree decomposition:

\begin{definition}
\label{def:lexSubtreeOrder}
Let $1,\dots,l$ be the children of $r$ in the tree $T$.
We define the \emph{lexicographical subtree order},
as the  order among the subtrees $(G_1,D_1),\dots,(G_l,D_l)$
which is given by:
$(G_i,D_i) < (G_j,D_j)$ iff there is 
a vertex $w \in V(G_i)\setminus X_r$ which has a smaller label than
every vertex in $V(G_j)\setminus X_r$.
\end{definition}

The algorithm non-deterministically guesses two main structures. 
First, we guess a tree decomposition of width $k$ for $H$.
This is done in a similar
way as in the \LogCFL\ algorithm from Wanke~\cite{Wa94} for testing that a graph
has bounded treewidth. 
Second, we guess an isomorphism $\phi$ 
from $G$ to $H$ by extending partial mappings from bag to bag.

Very simplified, Wanke's algorithm on input a graph $H$ starts  guessing a 
root bag and  it guesses then non-deterministically
further bags in the decomposition using the pushdown to test that   these bags 
fulfill the properties of a tree decomposition and 
that every edge in $G$ is included in some bag. Our algorithm simulates Wanke's algorithm as 
a subroutine. In the description of the new algorithm we concentrate on the isomorphism testing 
part and hide the 
details of how to choose the bags.
For simplicity 
the sentence ``guess a bag $X_j$ in $H$ according to Wanke's algorithm'' means that we
simulate the guessing steps from Wanke,  checking at the same time that the constructed structure is 
in fact a tree decomposition. Note, if the bags were not chosen appropriately, then 
the algorithm  would halt and reject.



We start 
guessing a root bag $X'_{r'}$ of size $\leq k+1$ for a decomposition of $H$.
With $X'_{r'}$ as root bag we guess the tree decomposition $D'$ of $H$ which corresponds
to $D$ and its root $r$.
We also construct a mapping $\phi$ describing
a partial isomorphism from the vertices of $G$ onto the vertices of $H$.
At the beginning, $\phi$ is the empty mapping and we guess an extension of $\phi$ from $X_r$ onto $X'_{r'}$.
The algorithm  starts with $a=r$ (and $a'=r'$).
Then  we describe isomorphism classes for $1,\dots,l$, the children of $a$.
First,
the children of $a$ can be distinguished
because $X_1,\dots,X_l$ may intersect with $X_a$ differently.
Second,
we further partition the children within 
one class according to the number of isomorphic siblings in that class.
This can be done in logspace with cross comparisons of pairs among $(G_1,D_1),\dots,(G_l,D_l)$, see Corollary~\ref{cor:TreeWIsoTwoDecompInL}.
It suffices to order the isomorphism classes according to the lexicographical subtree order of the 
members in the classes.
We compare then the children of $a$ with guessed children of $a'$ keeping the
following information:
For each isomorphism class we check whether there is the same number of isomorphic subtrees of $a'$ in $H$ and whether 
those intersect with $X'_{a'}$, accordingly.
For this we use the lexicographical subtree order to go through 
the isomorphic siblings from left to right, just keeping a pointer to the current child on the work tape.
For two such children, say $s_1$ of $a$ and $t_1$ of $a'$, 
we check then recursively whether $(G_1,D_1)$ is isomorphic to 
the corresponding subgraph of $t_1$ in $H$,
by an extension of  $\phi$.

When we go into recursion,  we push on the stack
$O(\log n )$ bits for a description of 
$X_a$ and $X'_{a'}$ as well as a description of the partial mapping  $\phi$ from
$X_a$ onto $X'_{a'}$.

In general, we do not keep all the information of $\phi$ on the stack.
We only have the partial isomorphism 
$\phi: \{ v \mid v \in X_r \cup \dots \cup X_a\} \rightarrow \{ v \mid v \in X'_{r'} \cup \dots \cup X'_{a'}\}$,
where $r,\dots,a$ ($r',\dots,a'$, respectively) is a simple path in $T$ from the root to the node at the current level of recursion.
After we ran through all children of some node  we go one level up in recursion
and recompute all the other information which is given implicitly by the subtrees from which we returned.
Suppose now, we returned to the bag $X_a$, we have to do the following:

\begin{itemize}
\item Pop from the stack the partial isomorphism $\phi$ of the bags $X_a$ onto $X'_{a'}$
\item Compute the lexicographical next isomorphic sibling.
	For this we consider the partition 
	into isomorphism classes according to $\phi$ 
	and the lexicographical subtree order of Definition~\ref{def:lexSubtreeOrder}.
	Recall, isomorphism testing of two subtrees of $X_a$ can be done in logspace.
\item If there is no such sibling then we 
	compute the lexicographical first child of $X_a$ inside the same isomorphism class.
	From this child of $X_a$ we compute the sibling which 
	is not in the same isomorphism class and which comes
	next to the right in the lexicographical subtree order.
\item If there is neither a further sibling in the same isomorphism class nor a non-isomorphic sibling of higher lexicographical order
	then we ran through all children of $X_a$ and we are ready to further return one level up in recursion.
\end{itemize}

Also for $X'_{a'}$ we guess all children in an isomorphism class from left to right in lexicographical subtree order.
If there is no further level to go up in recursion then the stack is empty and we halt in an accepting state.
Algorithm~\ref{alg:TreeWLogCFLIsoAlg} summarizes the above considerations.

\begin{algorithm}
\caption{Treewidth Isomorphism with one tree decomposition}
\label{alg:TreeWLogCFLIsoAlg}
\begin{flushleft}
{\bf Input:} Graphs $G,H$, tree decomposition $D$ for $G$, bags $X_a$ in $G$ and $X'_{a'}$ in $H$.\\
{\bf Top of Stack:} Partial isomorphism $\phi$ mapping the vertices in the parent bag of $X_a$ onto\\
\mbox{}\quad the vertices in the parent bag of $X'_{a'}$.\\
{\bf Output:} Accept, if $G$ is isomorphic to $H$ by an extension of $\phi$.
\\[-2ex]
\end{flushleft}
\begin{algorithmic}[1]
\STATE Guess an extension of $\phi$ to a partial isomorphism from $X_a$ onto $X'_{a'}$
\STATE {\bf if} $\phi$ cannot be extended to a partial isomorphism which maps $X_a$ onto $X'_{a'}$
	{\bf then} reject
\STATE Let $1,\dots,l$ be the children of $a$ in $T$.
	Partition the subtrees of $T$ rooted at $1,\dots,l$ into $p$ isomorphism classes $E_1,\dots,E_p$
\STATE {\bf for} each class $E_j$ from $j=1$ to $p$\\
\STATE \quad {\bf for} each subtree $T_i \in E_j$ (in lexicographical subtree order)\\
\STATE \qquad guess a bag $X'_{i'}$ in $H$ (in increasing lexicographical subtree order). Let $H_{i'}$ be the\\ 
	\mbox{}\qquad	subgraph of $H$ induced by the vertices in $X'_{i'}$ and by those which are separated\\ 
	\mbox{}\qquad	from $X'_{r'}$ in $H \setminus X'_{i'}$
\STATE \qquad {\bf if} $X'_{i'}$ is not a correct child bag of $X'_{a'}$ (see Wanke's algorithm) {\bf then} reject.
\STATE \qquad Invoke this algorithm with input $(G_i,H_{i'},D_i, X_i,X'_{i'})$ recursively and push $X_a$, $X'_{a'}$\\
	\mbox{}\qquad	and the partial isomorphism $\phi$ on the stack
\STATE \qquad After recursion pop these informations from the stack
\STATE {\bf if} the stack is not empty {\bf then} go one level up in recursion
\STATE accept and halt
\end{algorithmic}
\end{algorithm}


In Line~1, we guess an extension of 
 $\phi$ to include a mapping from $X_a$ onto $X'_{a'}$. 
We know the partial isomorphism of their parent bags since
this information can be found on the top of the stack.
In Line~3, 
we have e.g.\ the partition $E_1 = \{T_1, \dots, T_{l_1}\}$, $E_2 = \{T_{l_1+1},\dots,T_{l_2}\}$ and so on.
It 
can be obtained in logspace by testing isomorphism of 
the tree structures $(G_1,D_1),\dots,(G_l,D_l)$.
Two subtrees rooted at $X_i$ and $X_j$ are in the same isomorphism class 
iff
there is an automorphism in $G$  which maps 
$X_i$ onto $X_j$ and fixes their parent $X_a$ setwise.
In Lines~6 to~9, 
we guess $X'_{i'}$ in $H$ which corresponds to $X_i$, we test recursively whether
the corresponding subgraphs $G_i$ and $H_{i'}$ are isomorphic with an
extension of 
$\phi$.
In Line~7, we check whether $X'_{i'}$ fulfills the  properties of a correct tree-decomposition
as in Wanke's algorithm (i.e.\
$X'_{i'}$ must be a separating set which separates its split components from the vertices in $X'_{a'} \setminus X'_{i'}$).



To see that the algorithm correctly computes an isomorphism, we make the following observation.
A bag $X_a$ is a separating set which defines the connected subgraphs $G_1,\dots,G_l$.
These subgraphs do not contain the root $X_r$ and $V(G_i) \cap V(G_j) \subseteq X_a$ 
since we have a tree decomposition $D$ ($V(G_i)$ are the vertices of $G_i$).
We guess and keep from the partial isomorphism $\phi$ 
exactly those parts which correspond to the path from the roots $X_r$ and $X'_{r'}$ to the 
current bags $X_a$ and $X'_{a'}$.
Once we verified a partial isomorphism from one child component (e.g.\ $G_i$) of $X_a$ 
onto a child component (e.g.\ $H_{i'}$) of $X'_{a'}$,
for the other child components it suffices to know the partial mapping  of 
$\phi$ from $X_a$ onto $X'_{a'}$.

Observe that for each $v$ in $G$ in a computation path from the algorithm
there can only be a value for $\phi(v)$.
Clearly, if $G$ and $H$ are isomorphic then
the algorithm can guess the decomposition of $H$ which fits to $D$,
and the extensions of $\phi$ correctly.
In this case the NAuxPDA has some accepting computation.
On the other hand, if the input graphs are non-isomorphic then in every non-deterministic computation either the 
guessed tree decomposition of $H$ does not fulfill the conditions of a tree 
decomposition (and would be detected) or the partial isomorphism $\phi$ cannot
be extended at some point.
\qed 

Wanke's algorithm  decides in \LogCFL\ whether the treewidth of a graph is at most $k$ by guessing all 
possible tree decompositions. 
Using a result from~\cite{GLS} it follows that there is also a (functional) 
\LogCFL\ algorithm that
on input a bounded
treewidth graph computes a particular tree decomposition for it.
Since \LogCFL\ is closed under composition, from this result and
Theorem~\ref{thm:TreeWIsoOneDecompInLogCFL} we get:

\begin{corollary}
The isomorphism problem for  bounded treewidth graphs is in \LogCFL.
\end{corollary}

\paragraph{\bf Conclusions and open problems.}
We have shown that the isomorphism problem for graphs of bounded treewidth is
in the class \LogCFL\ and that 
isomorphism testing and canonization of
bounded tree distance width graphs is complete for \Log.
By using standard techniques in the area it can be shown that the 
same upper bounds apply for other problems related to isomorphism  on these graph classes. 
For example the automorphism problem
or the functional versions of automorphism and isomorphism
can be done within the same complexity classes. 
The main question remaining is whether the \LogCFL\ 
upper bound for isomorphism of bounded treewidth graphs can be improved.
On the one hand, no \LogCFL-hardness result for the isomorphism problem 
is known, 
so maybe the result can be improved. 
We believe
that proving a logspace upper bound for the isomorphism problem 
of bounded treewidth graphs would require to compute tree decompositions
within logarithmic space, which is a long standing open question.
Another interesting open question is whether 
bounded treewidth graphs can be canonized in \LogCFL.

\vspace{-1cm}

\end{document}